\documentclass[aps,prl,twocolumn,raggedbottom,superscriptaddress,showpacs]{revtex4}

\usepackage{graphicx}
\usepackage{dcolumn}
\usepackage{amsmath}
\usepackage{amssymb}
\usepackage{bm}

\newcommand{\2}{\frac{1}{2}}
\newcommand{\3}{\frac{3}{2}}

\begin{document}
\title{Kondo Screening in a Magnetically Frustrated Nanostructure:
Exact Results on a Stable, Non-Fermi-Liquid Phase}

\author{Kevin Ingersent}
\affiliation{Department of Physics, University of Florida,
Gainesville, FL 32611}
\author{Andreas W.\ W.\ Ludwig}
\affiliation{Department of Physics, University of California,
Santa Barbara, CA 93106}
\author{Ian Affleck}
\affiliation{Department of Physics and Astronomy, University of British
Columbia, Vancouver, B.C., Canada, V6T 1Z1}

\date{14 July 2005. Last revised: 7 December 2005}

\begin{abstract}
Triangular symmetry stabilizes a novel non-Fermi-liquid phase in the
three-impurity Kondo model with frustrating antiferromagnetic interactions
between half-integer impurity spins. The phase arises without fine-tuning of
couplings, and is stable against magnetic fields and particle-hole symmetry
breaking. We find a conformal field theory describing this phase, verify it
using the numerical renormalization group, and extract various exact,
universal low-energy properties. Signatures predicted in electrical transport
may be testable in scanning tunneling microscopy or quantum-dot experiments.
\end{abstract}


\maketitle

The same many-body physics that is responsible for the Kondo screening of
magnetic impurities in bulk metals \cite{Hewson:1993} produces resonances
in tunneling through a quantum dot \cite{dot-Kondo} or an adatom on
a metallic surface \cite{STM-Kondo}.
Greater experimental control over the latter settings
allows systematic study of multiple-``impurity'' configurations
in which the Kondo effect competes with ordering of the local degrees of
freedom \cite{Chen:1999,Jamneala:2001,double-dots}.
One goal that remains elusive is experimental realization of non-Fermi-liquid
(NFL) behavior similar to that exhibited, e.g., by the two-impurity
and two-channel Kondo models \cite{PRBwBJones,Cox:1998}.

A cluster of three antiferromagnetically coupled spins is of fundamental
importance as the simplest example of frustration, a feature of many
magnetic systems.
Scanning tunneling microscopy (STM) has found two distinct types of compact Cr
trimer on a gold surface \cite{Jamneala:2001}:
``type-2'' trimers show a sharp resonance of width (Kondo temperature)
$T_K\approx 50$\,K, whereas any Kondo effect for isolated Cr atoms
and ``type-1'' trimers seems to have $T_K\ll 7$\,K.
Attempts to explain this result via variational \cite{Kudasov:2002},
quantum Monte Carlo \cite{Savkin:2005}, and perturbative
renormalization-group (RG) \cite{Lazarovits:2005} treatments
of a three-impurity Kondo model have reached opposing conclusions
concerning the triangular geometry of the type-2 trimers:
equilateral \cite{Kudasov:2002,Lazarovits:2005}
or isosceles \cite{Savkin:2005}.
Interest is also developing in the interplay between Kondo physics and
interdot quantum entanglement in triangular quantum-dot devices
\cite{triple-dots}.

This Letter reports exact results on a frustrated phase of the three-impurity
Kondo model exhibiting NFL behavior distinctly different from that found in
previously studied models.
We present a conformal field theory (CFT), deduced by comparison with
numerical renormalization-group (NRG) results, showing that the NFL fixed point
is stable against particle-hole-symmetry breaking
(unlike its two-impurity counterpart), exchange anisotropy, and
even an applied magnetic field (which destroys two-channel Kondo NFL behavior).
This enhanced stability compared to other NFL fixed points makes the frustrated
phase an excellent candidate for experimental realizations; indeed, it has
been argued on the basis of weak-coupling RG \cite{Lazarovits:2005} to
describe the low-energy physics of the type-2 Cr trimers in
\cite{Jamneala:2001}.
We make predictions for the conductance expected in STM experiments on
trimers and in certain quantum-dot devices.

\textit{Model.}---We start with a Hamiltonian
$H_{\text{band}}+H_{\text{int}}$ describing
a noninteracting conduction band coupled via
\begin{equation}
\label{H_start}
H_{\text{int}} = J \sum_{j,\alpha,\beta}\psi^{\dagger,\alpha}(\vec{r}_j) \,
{\textstyle\2}\vec{\sigma}_{\alpha\beta} \, \psi_{\beta}(\vec{r}_j)
\cdot\vec{S}_j
\quad (J > 0)
\end{equation}
to spin-$S$ impurities $\vec{S}_j$ ($j=1,2,3$) at the vertices
$\vec{r}_j$ of an equilateral triangle;
$\psi_{\alpha}(\vec{r})$ annihilates an electron with spin $\alpha=\pm\2$
at $\vec{r}$.
We assume that the permutation group $S_3$ that maps the set $\{\vec{r}_j\}$
onto itself is a subgroup of the lattice symmetry group (as is the case,
e.g., in \cite{Jamneala:2001}).
The impurities couple to just six orthonormal combinations of conduction
states, annihilated by operators
$\psi_{h,\alpha}\propto \sum_j e^{-i2\pi j h/3}\psi_{\alpha}(\vec{r}_j)$,
where $h=0,\pm 1$ is the ``helicity'':
under a $2\pi/3$-rotation about the center of symmetry,
a helicity-$h$ state is multiplied by $e^{i 2 \pi h/3}$.
The combined states of the three impurities can also be constructed to have
well-defined helicities, in which case the Hamiltonian conserves total
helicity (modulo 3) and is invariant under interchange of all helicity labels
$1$ and $-1$.
Then,
Eq.~\eqref{H_start} can be rewritten \cite{Paul:1996}
\begin{eqnarray}
H_{\text{int}}
&=& \left[ J_{00}\,\vec{s}_{00}
   + J_{11}(\vec{s}_{11}+\vec{s}_{\bar{1}\bar{1}}) \right]
     \!\cdot\!{\vec{{\cal  S}}}_0
   + \left[ J_{01}(\vec{s}_{01}+\vec{s}_{\bar{1}0}) \right.
\nonumber \\
&+& \left. J_{1\bar{1}}\vec{s}_{1\bar{1}} \right]\!\cdot\!{\vec{{\cal  S}}}_1
    + \left[ J_{01}(\vec{s}_{10}+\vec{s}_{0\bar{1}})
    + J_{1\bar{1}}\vec{s}_{\bar{1}1} \right]
    \!\cdot\!{\vec{{\cal  S}}}_{\bar{1}} ,
\label{H_int}
\end{eqnarray}
where
${\vec{{\cal  S}}}_h = \sum_j e^{i2\pi j h/3} \vec{S}_j$,
$\vec{s}_{hh'}\!=\!\sum_{\alpha\beta} \psi^{\dagger,h,\alpha}
\2 \vec{\sigma}_{\alpha\beta} \psi_{h',\beta}$,
and $\bar{1}\equiv -1$;
$J_{hh'}$ equals $J$ times a non-negative factor that depends on the impurity
separation and the conduction-band dispersion, as well as $h$ and $h'$
\cite{ExplicitIsotropic}.

For $S=\2$, the NRG shows \cite{Paul:1996} that over a large region of the
parameter space of Eq.~\eqref{H_int},
the low-energy physics is governed by a ``frustrated'' fixed point at which
the impurities are locked into the
subspace of two doublets of combined spin $S_{\text{imp}}=\2$, one each of
helicity $h=\pm 1$.
Three spins of arbitrary half-integer $S$, coupled by an additional
Hamiltonian term $K\sum_{i<j}\vec{S}_i\cdot\vec{S}_j$ with $K\gg J$, also
lock into an $S_{\text{imp}}=\2$, $h=\pm 1$ subspace.
Weak-coupling RG analysis \cite{Lazarovits:2005} of this augmented model,
which for $S=\frac{5}{2}$ provides a description of equilateral Cr trimers,
is consistent with flow to the same fixed
point; for $S=\frac{5}{2}$, moreover, the characteristic temperature $T_K$
for this flow is found to greatly exceed the single-impurity Kondo scale,
in agreement with the Cr-trimer experiments \cite{Jamneala:2001}.

In the frustrated phase, $J_{1\bar{1}}$ in Eq.~\eqref{H_int} scales to zero,
$J_{00}$ and $J_{11}$ can be neglected, and particle-hole asymmetry is
marginal \cite{Paul:1996}.
Thus, we analyze the fixed point in a restricted $S_{\text{imp}}=\2$ space,
replacing Eq.~\eqref{H_int} by
\begin{equation}
\label{Hint}
H_{\text{int}} = -\sqrt{2} J_{01} [(\psi^\dagger {\textstyle\2}\vec{\sigma} \,
T^+ \psi) \tau_{\text{imp}}^- + \text{H.c.} ] \cdot
\vec{S}_{\text{imp}}.
\end{equation}
Here, $T^{\pm}$ and $T^z$ act on the electron helicity in the spin-1
representation of an ``orbital-spin'' $SU^{(t)}(2)$ \cite{Notation},
with matrix elements
$(T^z)_{h,h} = h$, $(T^{+})_{1, 0} = (T^{+})_{0,-1}=\sqrt{2}$.
The Pauli matrices $\vec{\sigma}$,
$2\vec{S}_{\text{imp}}\equiv 2\vec{\cal S}_0$ and
$\vec{\tau}_{\text{imp}}$ act, respectively, on the
electron spin, impurity spin, and impurity helicity,
with $(\tau_{\text{imp}}^z)_{h,h} = -h$ for $h=\pm 1$,
$(\tau_{\text{imp}}^+)_{-1,1} = 1$.

It is important to note that setting $J_{1\bar{1}}=0$ enlarges the
$S_3$ symmetry of Eq.~\eqref{H_int} to a $U^{(t)}(1)$ symmetry in
Eq.~\eqref{Hint}, replacing total helicity (conserved only modulo 3)
by a conserved quantity $t_z$: the eigenvalue of
$\psi^{\dagger}T^z\psi+\2\tau_{\text{imp}}^z$.
Now, $H_{\text{int}}$ commutes with $SU^{(s)}(2)$ spin, $U^{(t)}(1)$
orbital spin, and also with $SU^{(i)}(2)$ isospin defined by
$I^z = \2\sum_{h,\alpha}\;\psi^{\dagger,h,\alpha} \psi_{h,\alpha}$,
$I^+=\2\sum_{h,\alpha,\beta}\;\epsilon_{\alpha \beta}
\psi^{\dagger,h,\alpha} \psi^{\dagger,-h,\beta}$.

\textit{CFT description.}---We obtain exact analytical results for the
frustrated fixed point using the boundary CFT approach to quantum impurity
problems \cite{QuantumImpurities}.
The operator $\psi_{h,\alpha}$ is considered to act at the boundary
$x=0$ of a one-dimensional space $0\le x \le l$ \cite{PRBwBJones}.
The key is to find a ``conformal embedding '' (a decomposition of the
bulk fermions $\psi_{h,\alpha}(x)$ into products of ``constituent'' fields
\cite{ZamolodchikovFateev}) admitting a ``fusion procedure''
(generating a new conformally invariant boundary condition)
that reproduces the fixed-point finite-size spectrum (FSS).
We deduce this FSS by extending to higher accuracy the NRG results
of \cite{Paul:1996}.
From the fusion procedure, all universal low-energy properties
in the physical limit $l\rightarrow\infty$ can in principle be computed
exactly.

We first construct a conformal embedding of the free Dirac fermions
$\psi_{h,\alpha}(x)$ in which the helicities transform in the spin-1
representation of an $SU^{(p)}(2)$ ``pseudospin'' $\vec{P}$, where
$P^z=T^z$ and $P^+$ has matrix elements in the helicity basis
$(P^+)_{1,0}=-(T^+)_{1,0}$, $(P^+)_{0,-1}=(T^+)_{0,-1}$.
Unlike $\vec{T}$ defined above, $\vec{P}$ commutes with isospin $\vec{I}$.
The free-fermion FSS can be decomposed into products of
$SU^{(s)}(2)_3 \times SU^{(i)}(2)_3 \times SU^{(p)}(2)_8$
conformal towers \cite{Embedding}
as exemplified in Table~\ref{Table:free} for boundary conditions that
yield a nondegenerate ground state.
Here, $SU(2)_k$ is a level-$k$ Kac-Moody CFT;
see \cite{ZamolodchikovFateev} and references therein.

\begin{table}[t]
\renewcommand{\arraystretch}{1.2}
\caption{\label{Table:free}%
Finite-size spectrum of free fermions decomposed into products of spin,
isospin, and pseudospin conformal towers, labeled by
$s$, $i$, and $p$, respectively.
The subscript $\Delta$ gives each tower's contribution
to the excitation energy \protect\cite{Energies}.}
\begin{ruledtabular}
\begin{tabular}{llc}
$(s)_{\Delta}$ & $(i)_{\Delta}$ & $(p)_{\Delta}$ \\
\hline
$(0)_0$       & $(0)_0$       & $(0)_0 + (4)_2$ \\
$(\3)_{3/4}$  & $(\3)_{3/4}$  & $(0)_0 + (4)_2$ \\
$(\2)_{3/20}$ & $(\2)_{3/20}$ & $(1)_{1/5} + (3)_{6/5}$ \\
$(1)_{2/5}$   & $(1)_{2/5}$   & $(1)_{1/5} + (3)_{6/5}$ \\
$(0)_0$       & $(1)_{2/5}$   & $(2)_{3/5}$ \\
$(1)_{2/5}$   & $(0)_0 $      & $(2)_{3/5}$ \\
$(\2)_{3/20}$ & $(\3)_{3/4}$  & $(2)_{3/5}$ \\
$(\3)_{3/4}$  & $(\2)_{3/20}$ & $(2)_{3/5}$
\end{tabular}
\end{ruledtabular}
\end{table}

Since Eq.~\eqref{Hint} lowers the $SU^{(p)}(2)$ symmetry of
$H_{\text{band}}$ to $U^{(p)}(1) \equiv U^{(t)}(1)$, we analyze the
frustrated fixed point using an embedding obtained from that
above by decomposing $SU^{(p)}(2)_8 \supset U^{(t)}(1)_8 \times Z_8$,
where $Z_8$ is a parafermionic CFT \cite{ZamolodchikovFateev}.
In any $SU(2)_k$ CFT, each primary operator $\phi^{(j)}$, transforming
in the spin-$j$ representation ($j=0, \2, 1, ..., k/2$),
factors into a sum of products of a $Z_k$ primary $\psi^{j}_m$ and a
$U(1)_k$ boson exponential \cite{ZEightDetails}:
\begin{equation}
\label{embeddingZEight}
\phi^{(j)} = \sum_{j-m \in \mathbb{Z}} \;
\psi^{j}_m \, e^{i (m/\sqrt{k}) \varphi}.
\end{equation}
Setting $k=8$ and $j=p$, we rewrite the spectrum in Table~\ref{Table:free}
as products of
$SU^{(s)}(2)_3 \times SU^{(i)}(2)_3 \times U^{(t)}(1)_8 \times Z_8$
conformal towers.
These products also provide the operator spectrum for the noninteracting
model.

Using this conformal embedding, we are able to obtain the frustrated
fixed-point FSS from the free-fermion FSS by applying
a three-step fusion procedure:
(1) fusion with the $s=\3$ primary operator in $SU^{(s)}(2)_3$,
then
(2) fusion with the $p=\2$ primary operator in $SU^{(p)}(2)_8$,
then
(3) fusion with the $\psi^{p=0}_{m=2}$ primary operator in
$Z_8$ \cite{fusion}.

As an illustration, Table~\ref{Table:fss} lists all CFT states of energy
\cite{Energies} $E<1$ in the frustrated FSS for boundary conditions that
yield a degenerate free-fermion ground state \cite{FusionStepOne}, along
with energies of NRG levels having the same $(s,i,t_z)$ quantum numbers.
To fix the overall NRG energy scale, distorted by band discretization,
we match the lowest excitation of the \textit{free-fermion} spectrum to its
CFT counterpart.
Apart from small energy shifts (residual discretization effects),
the CFT and NRG spectra agree perfectly.
In fact, all 1810 CFT states with $E\leq 1.8$ have
been compared with and match NRG levels \cite{3ImpInPreparation}.

\begin{table}[t]
\renewcommand{\arraystretch}{1.4}
\caption{\label{Table:fss}%
Finite-size spectrum at the frustrated fixed point.
$U^{(t)}(1)$ and $Z_8$ conformal towers are labeled by
$t_z$ and $(p,m)$, respectively.
Each row represents a pair of states
related by a change in the signs of $t_z$ \textit{and} $m$.
$E$ is the CFT excitation energy and $E_{\text{NRG}}$ is the NRG
energy computed for a band discretization parameter $\Lambda=3$.
See also Table~\protect\ref{Table:free}.}
\begin{ruledtabular}
\begin{tabular}{llllll}
\multicolumn{1}{l}{$(s)_{\Delta}$} & \multicolumn{1}{l}{$(i)_{\Delta}$} &
\multicolumn{1}{l}{$(t_z)_{\Delta}$} & \multicolumn{1}{l}{$(p,m)_{\Delta}$} &
$E$ & $E_{\text{NRG}}$ \\
\hline
$(0)_0$ & $(0)_0$ &$(\3)_{9/32}$ &
$(\2,-\2)_{7/160}$ &$0$ & $0$ \\[.3ex]
\hline
$(\2)_{3/20}$ & $(\2)_{3/20}$  & $(\2 )_{1/32}$ &
$(\3,-\3)_{3/32}$ & $0.1$ & $0.1001$ \\[.3ex]
\hline
$(0)_0$ & $(1)_{2/5}$ & $(\2)_{1/32}$ & $(\3,-\3)_{3/32}$ &
$0.2$ &$0.2000$ \\
$(1)_{2/5}$ & $(0)_0$ & $(\2)_{1/32}$ & $(\3,-\3)_{3/32}$
&$0.2$ &$0.2000$ \\[.3ex]
\hline
$(\2)_{3/20}$ & $(\2)_{3/20}$ & $(\3)_{9/32}$ &
$(\2,-\2)_{7/160}$ & $0.3$ & $0.2996$ \\[.3ex]
\hline
$(0)_0$ & $(0)_0$ & $(\2 )_{1/32}$ & $(\2,-\3)_{127/160}$
&$0.5$ &$0.4968$ \\
$(0)_0$ & $(0)_0$ & $(\frac{5}{2})_{25/32}$  &
$(\2,+\2)_{7/160}$ & $0.5$ & $0.5020$ \\[.3ex]
\hline
$(\2)_{3/20}$ & $(\2)_{3/20}$ & $(\2)_{1/32}$ &
$(\3,+\frac{5}{2})_{19/32}$ & $0.6$ & $0.5971$ \\
$(\2)_{3/20}$ &$(\2)_{3/20}$ & $(\3)_{9/32}$ &
$(\3,-\2)_{11/32}$ & $0.6$ & $0.6040$ \\
$(1)_{2/5}$ & $(1)_{2/5}$ & $(\2 )_{1/32}$ &
$(\3,-\3)_{3/32}$ & $0.6$ & $0.6001$ \\[.3ex]
\hline
$(\2)_{3/20}$ & $(\3)_{3/4}$ & $(\2)_{1/32}$ &
$(\3,-\3)_{3/32}$ & $0.7$ & $0.7004$ \\
$(\3)_{3/4}$ & $(\2)_{3/20}$ & $(\2)_{1/32}$ &
$(\3,-\3)_{3/32}$ & $0.7$ & $0.7004$ \\
$(0)_0$ & $(1)_{2/5}$ & $(\2)_{1/32}$ &
$(\3,+\frac{5}{2})_{19/32}$ & $0.7$ & $0.7043$ \\
$(1)_{2/5}$ & $(0)_0$ & $(\2)_{1/32}$ & $(\3,+\frac{5}{2})_{19/32}$ &
$0.7$ & $0.7043$ \\
$(0)_0$ & $(1)_{2/5}$ & $(\3)_{1/32}$ &
$(\3,-\2)_{11/32}$ & $0.7$ & $0.6982$ \\
$(1)_{2/5}$  & $(0)_0$ & $(\3)_{1/32}$ &
$(\3,-\2)_{11/32}$ & $0.7$ & $0.6982$ \\[.3ex]
\hline
$(1)_{2/5}$ &$(1)_{2/5}$ &$(\3)_{9/32}$ &
$(\2,-\2)_{7/160}$ & $0.8$ & $0.8038$ \\
$(\2)_{3/20}$ & $(\2)_{3/20}$ & $(\2)_{1/32}$ &
$(\2,-\3)_{127/160}$ & $0.8$ & $0.8045$ \\
$(\2)_{3/20}$ &$(\2)_{3/20}$ & $(\frac{5}{2})_{25/32}$ &
$(\2,+\2)_{7/160}$ & $0.8$ & $0.8116$
\end{tabular}
\end{ruledtabular}
\end{table}

Applying the fusion procedure twice to the free-fermion FSS gives the complete
and exact spectrum of boundary operators that can be added to the fixed-point
Hamiltonian \cite{QuantumImpurities}. This spectrum (see Table~\ref{Table:op})
exhibits ``fractionalization'' of charge, spin, and orbital degrees of
freedom, as is typical of an NFL fixed point.
Remarkably, it also exhibits full $SU^{(p)}(2)$ symmetry. The
$SU^{(p)}(2)_8$ operator multiplets (last column of Table~\ref{Table:op})
should again be decomposed using Eq.~\eqref{embeddingZEight}
into $U^{(t)}(1)_8 \times Z_8$.

\begin{table}[b]
\renewcommand{\arraystretch}{1.2}
\caption{\label{Table:op}%
Operator spectrum at the frustrated fixed point.
$\Delta$ gives each factor's contribution to the scaling dimension.
``$2 \times$'' indicates two operators with the same $p$ and $\Delta$.}
\begin{ruledtabular}
\begin{tabular}{l@{\extracolsep{1.2em}}l@{\extracolsep{1em}}c}
$(s)_{\Delta}$ & $(i)_{\Delta}$ & $(p)_{\Delta}$ \\
\hline
$(0)_0$       & $(0)_0$       & $(0)_0 + (1)_{1/5} + (3)_{6/5}+ (4)_2$ \\
$(\3)_{3/4}$  & $(\3)_{3/4}$  & $(0)_0 + (1)_{1/5} + (3)_{6/5}+ (4)_2$ \\
$(\2)_{3/20}$ & $(\2)_{3/20}$ &
$(0)_0 + 2\!\times\![(1)_{1/5}\!+\!(2)_{3/5}\!+\!(3)_{6/5}] + (4)_2$ \\
$(1)_{2/5}$   & $(1)_{2/5}$   &
$(0)_0 + 2\!\times\![(1)_{1/5}\!+\!(2)_{3/5}\!+\!(3)_{6/5}] + (4)_2$ \\
$(0)_0$       & $(1)_{2/5}$   & $(1)_{1/5}+ 2\times(2)_{3/5}+ (3)_{6/5}$ \\
$(1)_{2/5}$   & $(0)_0$       & $(1)_{1/5}+ 2\times(2)_{3/5}+ (3)_{6/5}$ \\
$(\2)_{3/20}$ & $(\3)_{3/4}$  & $(1)_{1/5} + 2\times(2)_{3/5} + (3)_{6/5}$ \\
$(\3)_{3/4}$  & $(\2)_{3/20}$ & $(1)_{1/5} + 2\times(2)_{3/5} + (3)_{6/5}$
\end{tabular}
\end{ruledtabular}
\end{table}

Boundary operators entering the effective low-energy Hamiltonian for the
frustrated fixed point must respect the
$SU^{(i)}(2) \times SU^{(s)}(2) \times U^{(t)}(1)$ symmetry of the full
Hamiltonian~\eqref{Hint}.
Such operators appear in the first row of Table~\ref{Table:op}.
Only $(s, i, t_z, Z_8) = (0, 0, 0, (\psi^1_0)_{1/5})$ is relevant
(in the RG sense).
It cannot appear because it is odd under the $Z_2$ subgroup of $S_3$:
$\psi_{h,\alpha}\to -\psi_{-h,\alpha}$,
$\tau^-_{\text{imp}}\to \tau^+_{\text{imp}}$,
which is representable as a $\pi$-rotation about the $x$-axis
in orbital-spin space \cite{3ImpInPreparation}.
The least-irrelevant operator also respecting this discrete $Z_2$
symmetry of Eq.~\eqref{Hint} is the corresponding
$SU^{(p)}(2)_8$-descendant of dimension $\Delta=1+1/5$,
which yields a correction-to-scaling exponent
$1/5$ in excellent agreement with the value $0.200\pm
0.002$ observed in the NRG spectrum.

\textit{Physical results.}---We now present exact properties that can be
deduced from the CFT description.
Details, including analysis of the conditions required for observation
of these properties,
will appear elsewhere \cite{3ImpInPreparation}.

(a) \textit{Fixed-point properties.}---The frustrated fixed point has an
irrational ``ground-state degeneracy'' \cite{gTheorem}
$g=[\2(5\!+\!\sqrt{5})]^{1/2}$.
Moreover, in a quantum-dot device of triangular symmetry, where biases $V_j$
in leads $j=1,2,3$ produce in lead $i$ a current $I_i=\sum_j G_{ij} V_j$,
the $T=0$ zero-bias conductance is $G_{ii} = 4e^2/3h$.
By contrast, the ``isospin two-channel'' regime \cite{Paul:1996},
in which $J_{1\bar{1}}$ dominates Eq.~\eqref{H_int}, is unstable against
particle-hole asymmetry and at low energy exhibits the Fermi-liquid behavior
of the $SU(4)$ fixed point of [\onlinecite{triple-dots}(a)], with $g=1$ and
(in the limit of small particle-hole asymmetry) $G_{ii}=8e^2/9h < 4e^2/3h$.
The other stable fixed point of \cite{Paul:1996}, at which inter-impurity
correlations are irrelevant and the standard Kondo effect is recovered,
has $g=1$ and $G_{ii}=0$.

(b) \textit{Differential conductance.}---The leading irrelevant operator
of dimension $\Delta=1+1/5$ governs many properties near the fixed point.
In particular, the differential tunneling conductance into the impurities
from a metallic lead (e.g., an STM tip located symmetrically with
respect to the impurities) in the regime $k_B T, |e V| \ll k_B T_K$
($V$ being the bias voltage) has the form
$G_0^{-1} dI/dV \sim 1 - B (T/T_K)^{1/5} g[A e V/k_B T]$,
where $G_0$ is the $T=0$ linear-response conductance;
$A$ and $B$ are constants that can be fitted to experiment.
For $x\to 0$, $g[x] \to \text{const.}$, so
$G_0^{-1} dI/dV \sim 1 - B (T/T_K)^{1/5} g[0]$;
whereas $g[x] \sim c x^{1/5}$ (with $c$ a constant) for
$x \to \infty$, yielding
$G_0^{-1} dI/dV \sim 1  - cB(A eV / k_B T_K)^{1/5}$
\cite{FanoEffect}.
To lowest (quadratic) order in the tunneling matrix element between the
impurities and the lead \cite{GlazmanPustilnik}, the universal scaling
function $g[x]$ equals the exact function given in
[\onlinecite{QuantumImpurities}(b)].
Similar (and in linear response, identical) behavior is expected in
transport through triangular quantum-dot devices
\cite{3ImpInPreparation}.

(c) \textit{Breaking of particle-hole symmetry.}---This
lowers the isospin $SU(2)$ symmetry to the
$U(1)$ subgroup that conserves global charge
$2I^z$, while preserving the discrete $S_3$ symmetry.
The spectrum in Table~\ref{Table:op} is reclassified
by applying Eq.~\eqref{embeddingZEight}
to $SU^{(i)}(2)_3 \supset U^{(i)}(1) \times Z_3$.
The most-relevant operators that become allowed in the
low-energy Hamiltonian are marginal:
the charge current operator $2 I^z$, which is
exactly marginal and corresponds to a simple phase shift
\cite{QuantumImpurities}; and a degenerate pair
$(s, I^z, Z_3, t_z, Z_8)=
((0)_0, (0)_0, (\psi^1_0)_{2/5}, (0)_0, (\psi^2_4)_{3/5})$
arising from Table~\ref{Table:op}, row 5.
The last two operators are the boundary limits
of the left- and right-moving bulk currents
$J_{L,R} = \psi_{L,R}^\dagger[(T^z)^2 - \frac{2}{3} \mathbf{1}]\psi_{L,R}$
(Table~\ref{Table:free}, row 5) \cite{LR}. $J_{L,R}$
generate a $U(1)$ symmetry of the free-fermion bulk theory
\textit{not} preserved by the boundary condition.
The boundary limits of such operators are exactly marginal
\cite{3ImpInPreparation}, consistent with NRG results in the presence
of particle-hole asymmetry \cite{Paul:1996}.
Like a phase shift, the three exactly marginal deformations of the boundary
conditions affect the FSS (and the boundary limits of $J_{L,R}$ affect
the $T=0$ zero-bias conductance), but not the operator spectrum in
Table~\ref{Table:op} \cite{3ImpInPreparation}.
Thus, the NFL fixed point and its signatures, including the ground-state
degeneracy and power laws in the conductance, persist away from
particle-hole symmetry (unlike, e.g., the
NFL behavior of the two-impurity Kondo model \cite{PRBwBJones}).

(d) \textit{Other symmetry-breaking perturbations.}---It can be deduced
from Table~\ref{Table:op} that
(i) spin-orbit coupling is relevant, with dimension 3/5,
(ii) breaking of $S_3$ symmetry (e.g.,
through distortion of the equilateral triangular impurity geometry)
is relevant with dimension 1/5 \cite{Isosceles},
(iii) spin-exchange anisotropy is irrelevant,
(iv) a Zeeman field acting only on the impurity spins is exactly marginal,
and (v) the coupling $J_{1{\bar 1}}$ in Eq.~(\ref{H_int}) is irrelevant.
The implications of these results will be discussed elsewhere
\cite{3ImpInPreparation}.

In summary, we have found the exact low-energy behavior of a non-Fermi-liquid
phase arising from the interplay of magnetic frustration and Kondo physics
in the three-impurity Kondo model. The phase is stable against particle-hole
asymmetry, exchange anisotropy, and magnetic fields. It should be detectable in
tunneling into magnetic adatoms on metallic surfaces and in electrical
transport through triangular quantum-dot devices.

We are grateful for discussions with M.\ Fabrizio, D.\ Seo, and G.\ Zar\'{a}nd,
and for the hospitality of the Max-Planck-Institut MPIPKS (Dresden)
and the KITP (Santa Barbara), where portions of this work were performed.
This work was supported in part by NSF Grants No.\ PHY-990794 (K.I., I.A.),
No.\ DMR-0075064 (A.W.W.L.), and No.\ DMR-0312939 (K.I.), by NSERC (I.A.), and
by the Canadian Institute for Advanced Research (I.A.).


\begin{thebibliography}{28}
\bibitem{Notation}
Each Lie group is labeled with a superscript ($s$, $i$, $t$, or $p$)
denoting the subspace in which the group acts.
\bibitem{Energies}
All FSS energies are expressed as multiples of $\pi v_F/l$,
where $v_F$ is the Fermi velocity.
\bibitem{ZEightDetails}
In Eq.~\protect\eqref{embeddingZEight}, $m$ and $m \pm k$ are identified, so
$m$ (where $j-m \in \mathbb{Z}$) can be chosen such that
$-(k-1)/2 \leq m \leq k/2$.
Also, operators $\psi^{j}_m$ and $\psi^{k/2-j}_{-(k/2 -m)}$ are identified.
\bibitem{FanoEffect}
Possible Fano interference with direct tunneling from the STM to the
substrate will be discussed in \protect\cite{3ImpInPreparation}.
\bibitem{LR}
See~\cite{QuantumImpurities} concerning the operator spectra of the free
left-moving field $\psi_L$ and its right-moving counterpart $\psi_R$.
\bibitem{Isosceles}
If isosceles triangular symmetry and particle-hole symmetry are
preserved, there is just one relevant operator.
\end{thebibliography}
\end{document}